\documentclass{article}

\usepackage{placeins}

\usepackage[english]{babel}

\usepackage[letterpaper,top=2cm,bottom=2cm,left=3cm,right=3cm,marginparwidth=1.75cm]{geometry}

\usepackage{amsmath}
\usepackage{graphicx}
\usepackage[colorlinks=true, allcolors=blue]{hyperref}

\title{Mud-Standoff Effect Correction Based on Open-Short Calibration and Resistivity Consistency-Constrained Iterative Inversion for Oil-Based Mud Imagers}
\author{Li Fengfeng (email: cj21701@163.com), Feng Zhou, Wu Hongliang, \\
Zhang Hao, Tian Han, Liu Peng and Yuan Lixin\\
\\
\small PetroChina Research Institute of Petroleum Exploration and Development, Beijing 100083, China
}

\begin{document}
\maketitle

\begin{abstract}
We propose a mud-standoff effect correction method and a set of approximate apparent resistivity inversion methods suitable for oil-based mud micro-resistivity imaging logging. To calibrate the influence of the mud layer on electrode measurement signals, this study integrates the Open-Short calibration(OSC) method with the three-layer impedance model of the oil-based mud resistivity imager. By treating the electrode and the mud layer as an integrated whole and simulating the open/short-circuit states via the finite element method, the independent extraction of the mud layer impedance signal is achieved, and the formation impedance signal is separated from the total impedance. For fast inversion of formation resistivity, standoff thickness (mud layer thickness), and relative permittivity of formation, a resistivity consistency-constrained iterative inversion method is further proposed. In this method, the formation impedance is first converted into resistivity, and then the consistency residual of the resistivity at different frequencies is used as the objective function to invert the approximate apparent resistivity of the formation through an iterative optimization algorithm. The effectiveness of the finite element-simulated OSC method, the objective function construction scheme, and the consistency inversion method is verified through both numerical models and an example of field data.
\end{abstract}

\section{Introduction}

With the continuous advancement of oil and gas exploration towards complex domains such as deep formations, high-temperature and high-pressure(HTHP)  reservoirs, and shale reservoirs, oil-based mud (OBM) has been widely applied due to its superior lubricity, stability, and adaptability to HTHP conditions. However, the non-conductive nature of OBM makes traditional electrode-type resistivity logging methods inapplicable directly \cite{1}, \cite{2}. Compared with water-based mud (WBM) electrical imaging logging, OBM electrical imaging logging based on the capacitive coupling principle employs MHz-level high-frequency signals to drive current through the non-conductive mud and return to the return electrodes. Combined with the multi-frequency measurement mode, it significantly improves the resolution capability for formation resistivity and permittivity \cite{3}, \cite{4}, \cite{5}.

The raw measurement signals of such logging tools are susceptible to interference from mud resistivity, permittivity, and standoff, failing to directly and accurately characterize the true formation information. Currently, techniques such as orthogonal projection, optimization inversion, and machine learning are commonly used to decouple the real and imaginary parts of the raw complex impedance, constructing resistivity-dominated images, permittivity-dominated images, and standoff images, respectively, to effectively separate mud effects from formation properties \cite{6}, \cite{7}, \cite{8}. Among these methods, the result obtained by orthogonal projection is the stitched formation impedance, whose accuracy largely depends on the calculation precision of the mud impedance phase angle; the vertical/parallel approximation method can achieve rapid calculation of approximate apparent resistivity for formations within a wide dynamic range by calculating apparent resistivity for low-, medium- and high-resistivity formations separately. However, there is still a certain deviation between the equivalent mud impedance and the actual mud impedance response. Additionally, the final apparent resistivity curve needs to be stitched using segmental thresholds, which introduces inherent uncertainties \cite{9}, \cite{10}; although optimization inversion or neural network-driven quantitative inversion methods can achieve higher-precision formation parameter calculation, they have obvious limitations: the former requires pre-construction of large-scale training datasets and exhibits low inversion efficiency, making it difficult to meet the demand for rapid on-site processing; the latter suffers from poor result accuracy and weak controllability of the model prediction process\cite{11}, \cite{12}.

Open-Short Calibration (OSC) is an impedance calibration technique based on dual extreme state measurements. By acquiring the system's responses under two ideal boundary conditions (complete open circuit and complete short circuit), an error model is constructed to eliminate the deviation introduced by the measurement system itself, thus enabling the accurate extraction of the true impedance of the device under test \cite{13}. In this paper, this technique is extended to correct the interference of the mud layer on electrode measurement signals. We establish a three-layer impedance model ("electrode-mud layer-formation") that integrates the measuring electrode and mud layer as a single system component. Using the finite element method (FEM), we simulate the impedance responses under open- and short-circuit conditions. Based on these simulations, we construct an efficient approximate model through polynomial fitting. Ultimately, we apply the OSC formula to remove mud layer interference and extract the formation impedance.

\begin{figure}
    \centering
    \includegraphics[width=0.5\linewidth]{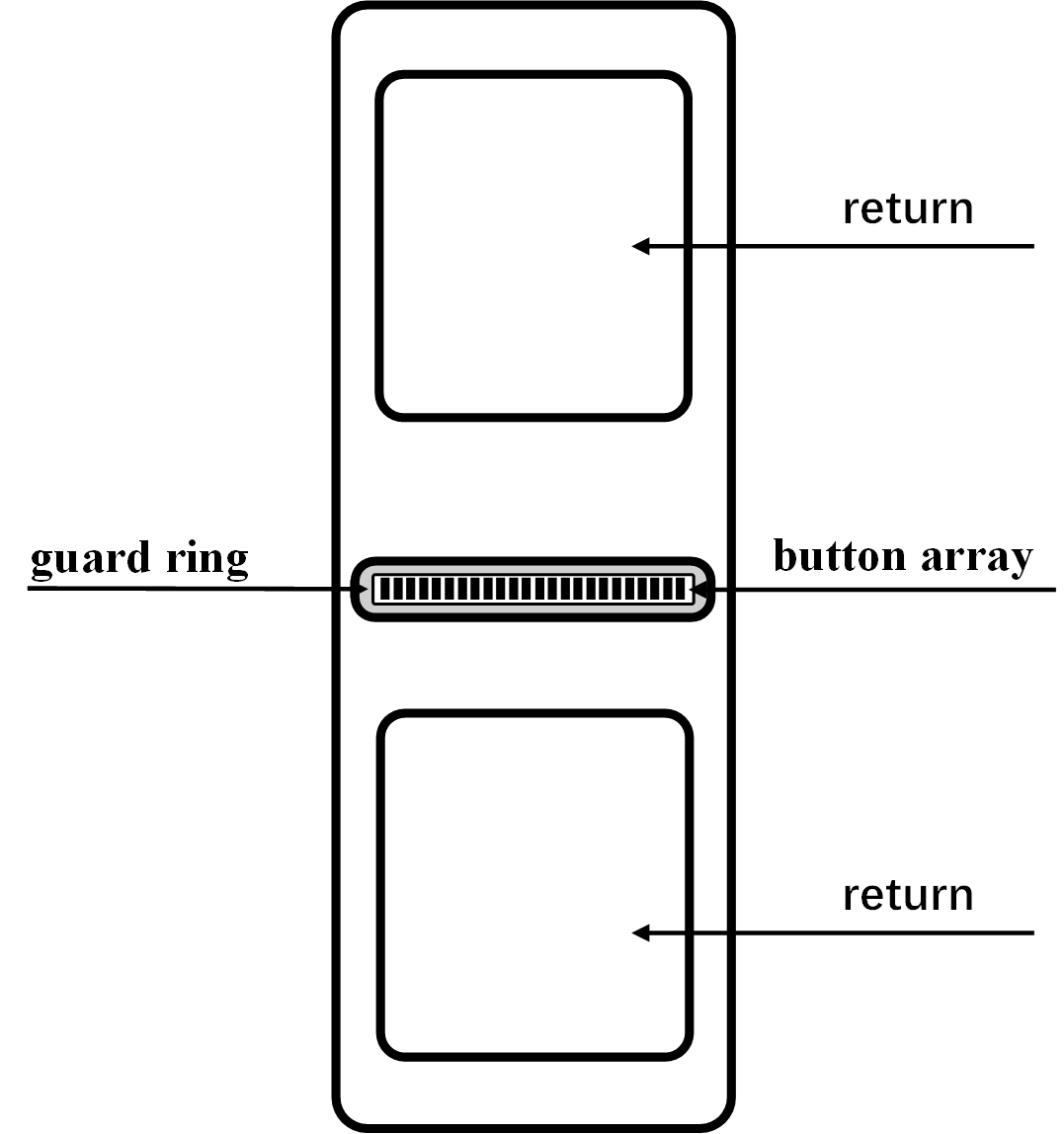}
    \caption{Schematic of fracture model and imaging pads}
    \label{fig:Pads}
\end{figure}

Theoretically, after converting the complex impedance into resistivity and permittivity, the formation resistivity values at different frequencies should be consistent. Based on this, this paper proposes to use the consistency residual of apparent resistivity at multiple frequencies as the objective function for formation resistivity inversion. The objective function and the corresponding inversion method have been verified through numerical models and field well data.
The pad structure of the imaging logging tool adopted in this study is shown in Fig.\ref{fig:Pads}, which includes 24 button electrodes, 1 annular focusing electrode, and 2 return electrodes, with two operating frequency modes: F1 (low frequency) and F2 (high frequency).

\section{Calculation Methods}

\subsection{Apparent resistivity and permittivity}

The total impedance recorded by the electrode is defined as:
\begin{equation}
    Z=\frac{1}{\sigma+i\omega\varepsilon\varepsilon_0}=\frac{\sigma}{\sigma^2+\omega^2(\varepsilon\varepsilon_0)^2}-i\frac{\omega\varepsilon\varepsilon_0}{\sigma^2+\omega^2(\varepsilon\varepsilon_0)^2}
\end{equation}
where $\sigma$ is the conductivity $\mathrm{(S/m)}$; $\varepsilon$ is the relative permittivity; $\varepsilon_0$ is the permittivity of free space $\mathrm{(F/m)}$; $\omega$ is the angular frequency $\mathrm{(rad/s)}$.

Let the real part of the impedance be $A=\frac{\sigma}{\sigma^2+\omega^2(\varepsilon\varepsilon_0)^2}$ and the imaginary part be $B=-\frac{\omega\varepsilon\varepsilon_0}{\sigma^2+\omega^2(\varepsilon\varepsilon_0)^2}$. The ratio of the imaginary part to the real part is expressed as:
\begin{equation}
    \frac{B}{A}=-\frac{\omega\varepsilon\varepsilon_0}{\sigma}
\end{equation}

Substituting Equation (2) into the expression for the real part of the impedance, we obtain:
\begin{equation}
    A=\frac{\sigma}{\sigma^2+\omega^2\left(\frac{\sigma B}{\omega A}\right)^2}=\frac{1}{\sigma\left(1+\frac{B^2}{A^2}\right)}
\end{equation}

From this, the apparent conductivity is derived as:
\begin{equation}
    \sigma_a=\frac{A}{A^2+B^2}
\end{equation}

Substituting Equation (4) into Equation (2), the apparent relative permittivity is further obtained:
\begin{equation}
    \varepsilon_a=-\frac{B}{\omega\left(A^2+B^2\right)\varepsilon_0}
\end{equation}

The total impedance measured by the electrode can be expressed as:
\begin{equation}
    Z_{totle}=K\frac{U}{I}=A+iB
\end{equation}
where $K$ is the instrument constant; $U$ is the potential of the button electrode; $I$ is the current of the button electrode.

Combining Equations (4), (5), and (6), the apparent resistivity $R_a$ and apparent relative permittivity $\varepsilon_a$  are finally expressed as:
\begin{equation}
    R_a=\frac{A^2+B^2}{A}
\end{equation}

\begin{equation}
    \varepsilon_a=-\frac{B}{\omega\left(A^2+B^2\right)\varepsilon_0}
\end{equation}

\subsection{Open-short calibration for standoff effect correction}
The equivalent circuit of electrical imaging logging can be simplified to a series impedance model of the formation and mud layer ($Z_{total} = Z_f + Z_m$). If the accurate mud impedance $Z_m$ can be obtained, the formation impedance $Z_f$ can be derived by subtracting the mud impedance from the total impedance.

1)  Equivalent mud impedance

In electrical imaging logging, the electrode is closely attached to the borehole wall, and the current flows vertically from the electrode through the mud into the formation. As a thin layer between the electrode and the formation, the mud can be equivalent to a parallel-plate model, and its impedance is calculated by a series model of parallel-plate capacitance and resistance \cite{7}:
\begin{equation}
    Z_m=\frac{d_m}{S}\left(\frac{\sigma_m}{{\sigma_m}^2+\omega^2(\varepsilon_m\varepsilon_0)^2}-i\frac{\omega\varepsilon_m\varepsilon_0}{{\sigma_m}^2+\omega^2(\varepsilon_m\varepsilon_0)^2}\right)
\end{equation}

The formation impedance is obtained by subtracting the mud impedance from the total impedance:
\begin{equation}
    Z_f = Z_{\text{total}} - K \cdot Z_m = A + \mathrm{i}B
\end{equation}

Substitute the formation impedance $Z_f$ into Equations (7) and (8) to calculate the formation apparent resistivity. It should be noted that there is a certain error between the mud impedance calculated by the equivalent resistance and capacitance and the actual mud impedance, but it can meet the requirements for approximate apparent resistivity calculation in high-resistivity formations \cite{7}.

2)  Standoff effect correction based on finite element method and OSC method

To accurately separate and calibrate the influence of the mud layer on the measured impedance signal, the Open-Short Calibration (OSC) method is adopted. The core idea of this method is: under the same geometric structure, measure the impedance responses under two extreme boundary conditions (open circuit and short circuit) respectively to construct an equivalent circuit model, and then separate the impedance contribution of the mud from the measured total impedance.

\begin{figure}[!htbp]
    \centering
    \includegraphics[width=0.5\linewidth]{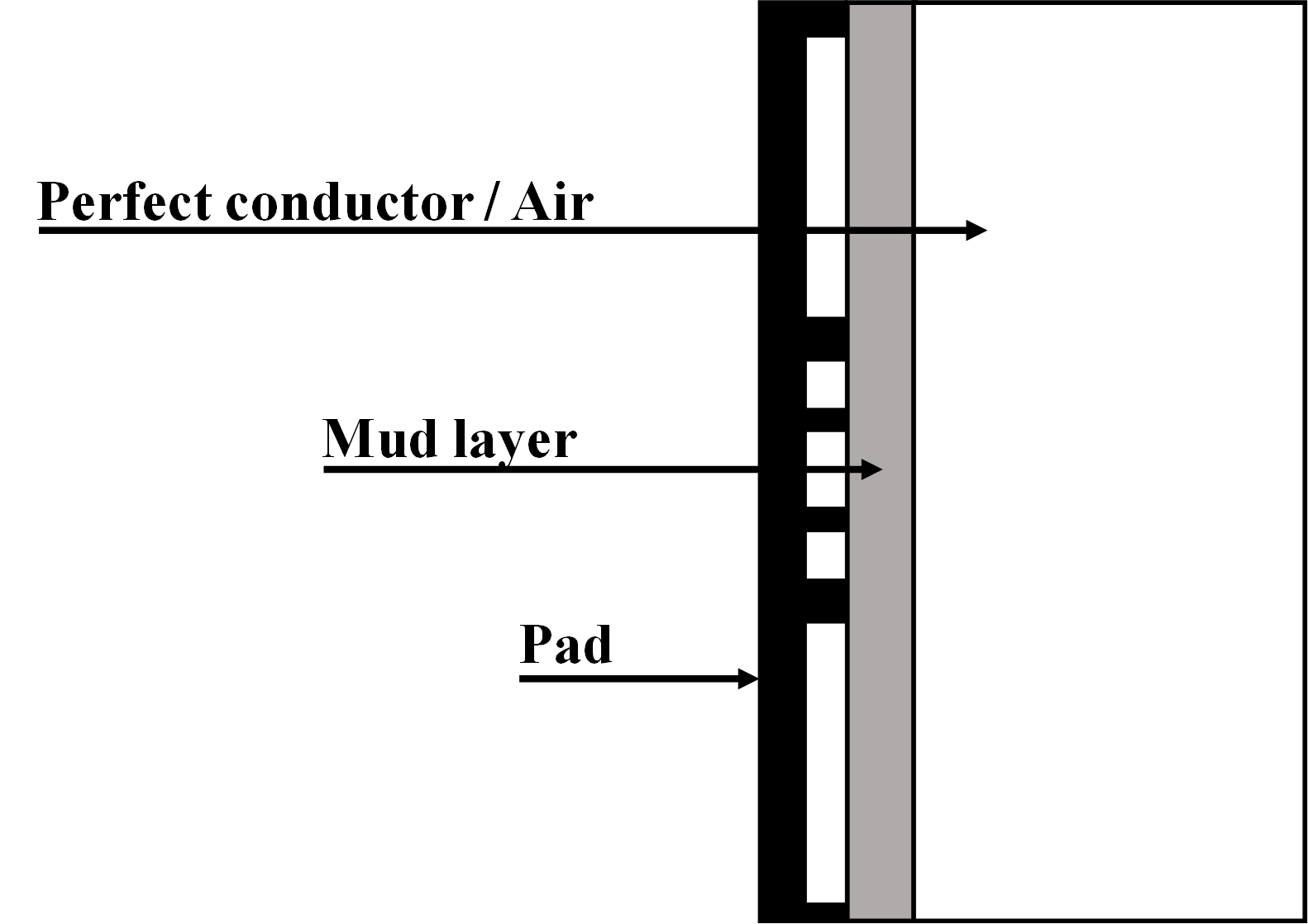}
    \caption{The calculation model of OSC}
    \label{fig:2}
\end{figure}

As shown in Fig.\ref{fig:2}, the calibration model consists of three layers: the outermost layer is an ideal conductor (short-circuit state) or air (open-circuit state), the middle layer is the mud to be calibrated, and the inner layer is the measuring pad. Let the total measured impedance be $Z_{total}$, the mud impedance be $Z_m$, the pad impedance be $Z_b$, and the external medium impedance be $Z_{ext}$.

In the short-circuit state, the external medium is an ideal conductor with approaches zero, and the total impedance of the measurement system is approximately:
\begin{equation}
    Z_{short}=Z_m+Z_b
\end{equation}

In the open-circuit state, the external medium is air with approaches infinity, and the total impedance of the measurement system is approximately:
\begin{equation}
    Z_{open}=Z_{air}+Z_m+Z_b
\end{equation}

The open-circuit state provides the maximum external impedance reference. Taking the open-circuit state as the "zero-signal" reference, the effective admittance change is defined as:
\begin{equation}
    Y_{eff}=\frac{1}{Z_{total}-Z_{short}}-\frac{1}{Z_{open}-Z_{short}}
\end{equation}
where $Z_{total}-Z_{short}=Z_b+Z_m$, $Z_{open}-Z_{short}=Z_{air}$

The external medium impedance can be obtained through short-circuit calibration. However, when the formation resistivity is relatively high, retaining the second term for differential calibration can further improve the calculation accuracy. At this time, the formation impedance after open-short calibration is expressed as:
\begin{equation}
    Z_f=Z_C=\left[\frac{1}{Z_{total}-Z_{short}}-\frac{1}{Z_{open}-Z_{short}}\right]^{-1}
\end{equation}

\begin{figure}[!htbp]
    \centering
    \includegraphics[width=1\linewidth]{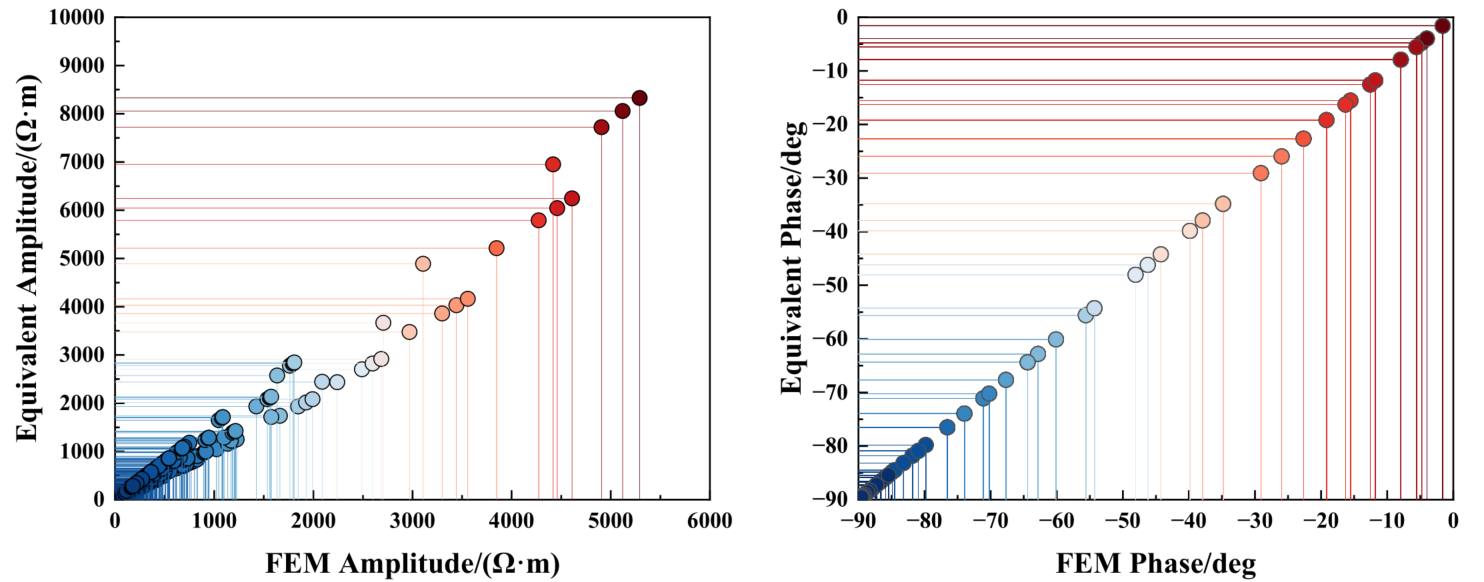}
    \caption{Comparison of mud impedance amplitude and phase}
    \label{fig:3}
\end{figure}

\begin{figure}[!htbp]
    \centering
    \includegraphics[width=1\linewidth]{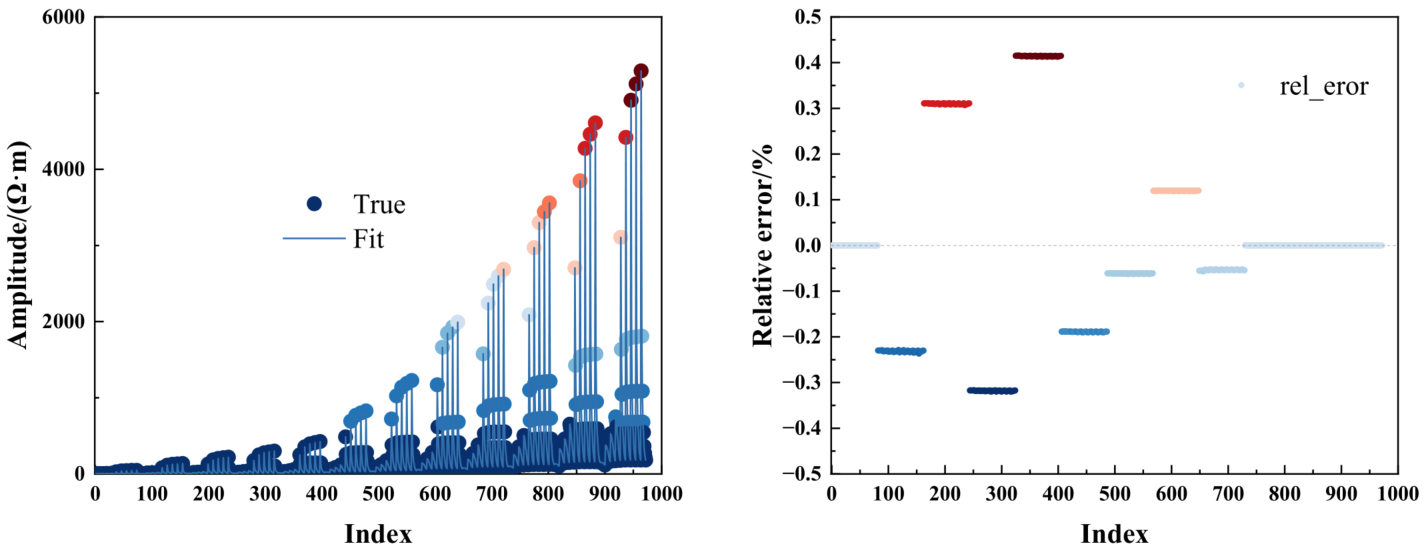}
    \caption{The fitting results of mud impedance amplitude}
    \label{fig:4}
\end{figure}

The finite element method is used to simulate the measured impedance under open-circuit and short-circuit states. Fig.\ref{fig:3} shows the comparison results between the measured impedance amplitude, phase angle calculated in the short-circuit state and the equivalent mud impedance: the phase angles are consistent, but the amplitudes differ significantly. A fourth-order polynomial (Eq.15) is used for piecewise fitting of the equivalent mud impedance amplitude and the finite element-calculated mud impedance amplitude. The fitting results, as shown in Fig.\ref{fig:4}, have a relative error of less than 0.5\% between the fitted values and the true values.

\begin{figure}[!htbp]
    \centering
    \includegraphics[width=0.5\linewidth]{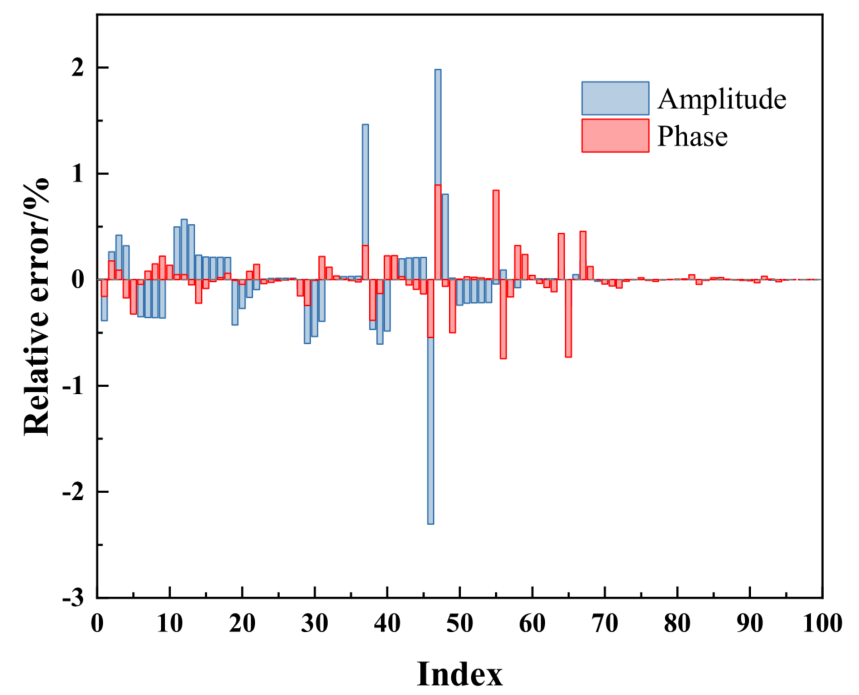}
    \caption{The relative error of impedance and phase in the open state}
    \label{fig:5}
\end{figure}

Similarly, a fourth-order polynomial is used to fit the measured impedance in the open-circuit state. The relative errors of the fitted impedance amplitude and phase are shown in Fig.\ref{fig:5}, with the maximum relative error not exceeding 2.5\%. Subsequently, the efficient calculation of impedance under open-circuit and short-circuit states will be achieved using the fitting results and the equivalent mud impedance phase.
\begin{equation}
    \begin{aligned}
A_{\text {True }} & =c_{0}+ \\
& c_{1} A+c_{2} A^{2}+c_{3} A^{3}+c_{4} A^{4}+ \\
& c_{5} d_{m}+c_{6} d_{m}^{2}+c_{7} d_{m}^{3}+c_{8} d_{m}^{4}+ \\
& c_{9} d_{m} A+c_{10} d_{m} A^{2}+c_{11} d_{m} A^{3}+ \\
& c_{12} d_{m}^{2} A+c_{13} d_{m}^{2} A^{2}+c_{14} d_{m}^{3} A
\end{aligned}
\end{equation}

\subsection{Objective function of consistency inversion}

Under the same formation resistivity condition, the impedance measured by the imaging tool at different frequencies shows significant frequency-dependent differences due to the influence of the borehole environment, dielectric constant, and other factors. 
Considering that the relative permittivity of the formation has frequency dispersion characteristics while the resistivity is non-dispersive, there should be a unique solution when inverting formation resistivity from multi-frequency impedance data. We construct an optimization objective function constrained by the consistency of apparent resistivity, and realize the iterative inversion of formation apparent resistivity by minimizing the residual of apparent resistivity between different frequencies.

The objective function is constructed as the residual of apparent resistivity at each frequency:
\begin{equation}
    F=\left\|\frac{R_{a, 1}-R_{a, 2}}{R_{a, 2}}\right\|
\end{equation}

In the inversion process, 5 parameters need to be iteratively optimized: the mud resistivity at two frequencies, the mud relative permittivity at two frequencies, and the mud thickness. The calculation process is as follows:

1)	Set the initial values and search ranges of mud resistivity, permittivity, and standoff. 
2)	Calculate the mud impedance and the approximate apparent resistivity of the formation at different frequencies based on the initial values; 

3)	Substitute the approximate apparent resistivity into Eq. 13 to compute the objective function value; 

4)	Iterate via the optimization algorithm until the objective function value meets the allowable precision.

\section{Methods validation}

\subsection{Validation of the OSC method for standoff effect correction}

\begin{figure}[!htbp]
    \centering
    \includegraphics[width=0.8\linewidth]{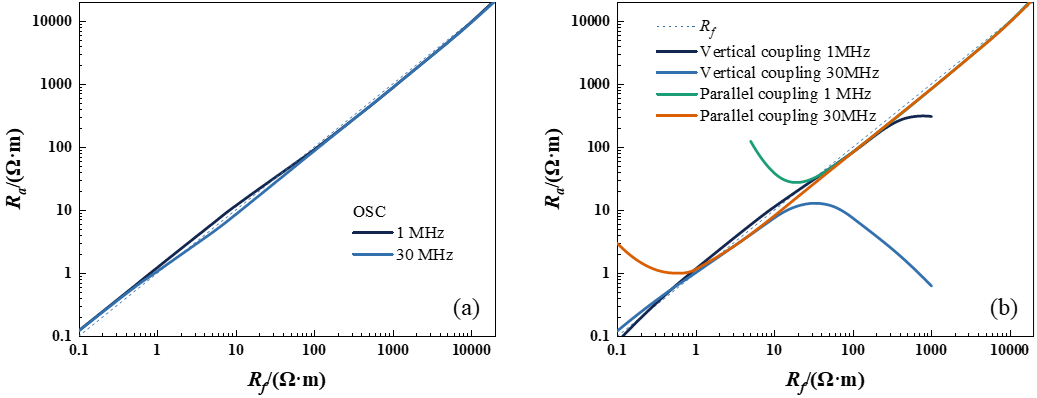}
    \caption{OSC corrected and vertical/parallel coupling results ($R_m$ = 10000 $\Omega \cdot \mathrm{m}$, $\varepsilon_m$ = 8, $stdf$ = 1 mm, $\varepsilon_f$ = 20)}
    \label{fig:6}
\end{figure}

\begin{figure}[!htbp]
    \centering
    \includegraphics[width=0.8\linewidth]{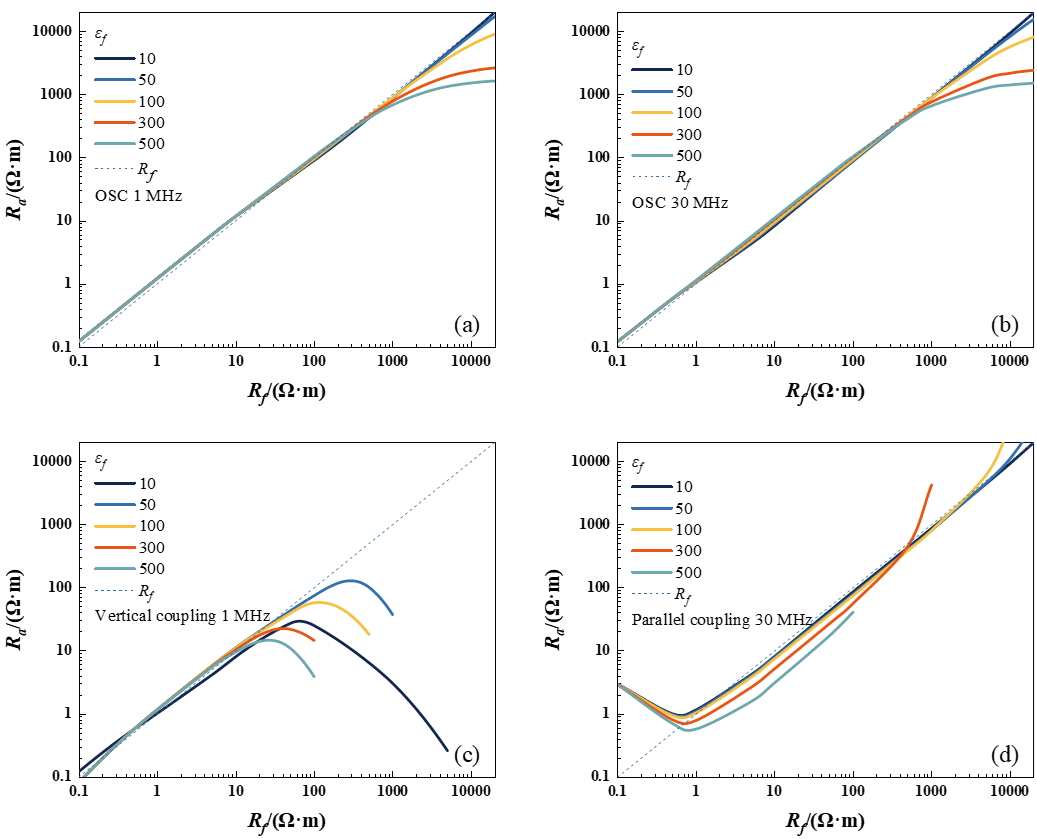}
    \caption{Apparent resistivity results under different formation permittivities ($R_m$ = 10000 $\Omega \cdot \mathrm{m}$, $\varepsilon_m$ = 8, $stdf$ = 1 mm, $\varepsilon_f$ = 20). (a) OSC corrected at 1 MHz. (b) OSC corrected at 30 MHz. (c) Vertical Approximation at 1 MHz. (d) Parallel Approximation at 30 MHz}
    \label{fig:7}
\end{figure}

\begin{figure}[!htbp]
    \centering
    \includegraphics[width=0.8\linewidth]{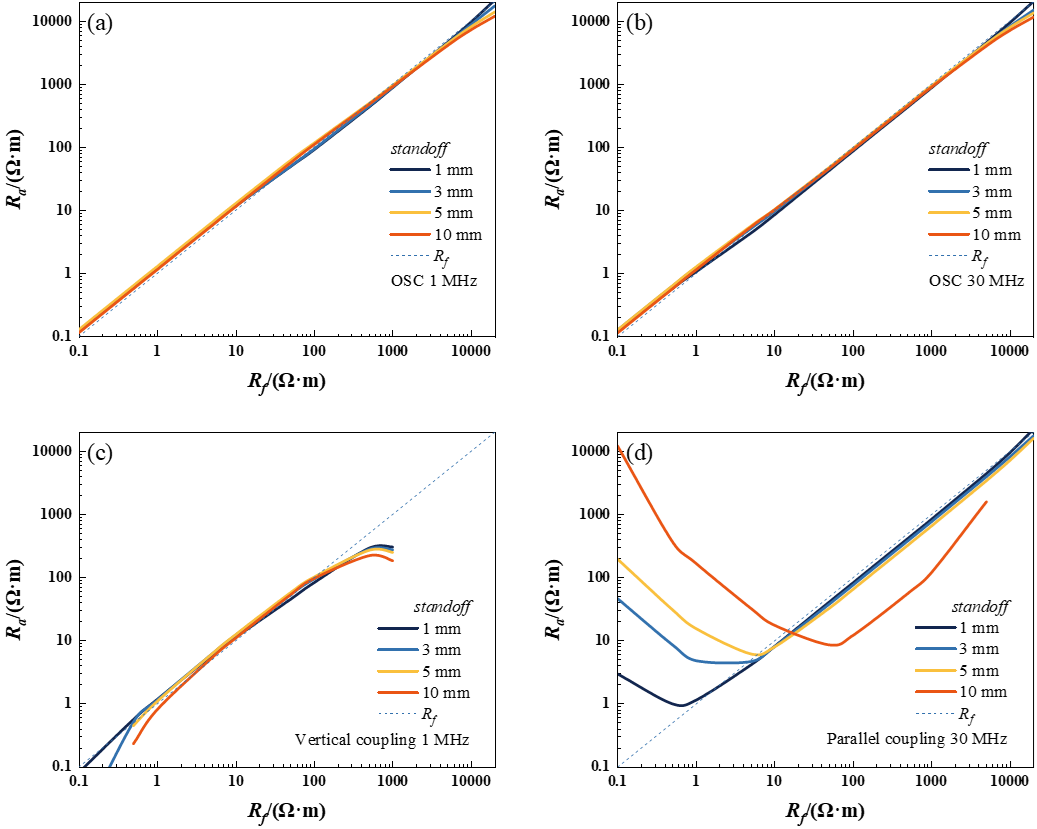}
    \caption{Apparent resistivity results under different standoffs ($R_m$ = 10000 $\Omega \cdot \mathrm{m}$, $\varepsilon_m$ = 8, $\varepsilon_f$ = 20). (a) OSC corrected at 1 MHz. (b) OSC corrected at 30 MHz. (c) Vertical Approximation at 1 MHz. (d) Parallel Approximation at 30 MHz}
    \label{fig:8}
\end{figure}

\begin{figure}[!htbp]
    \centering
    \includegraphics[width=0.8\linewidth]{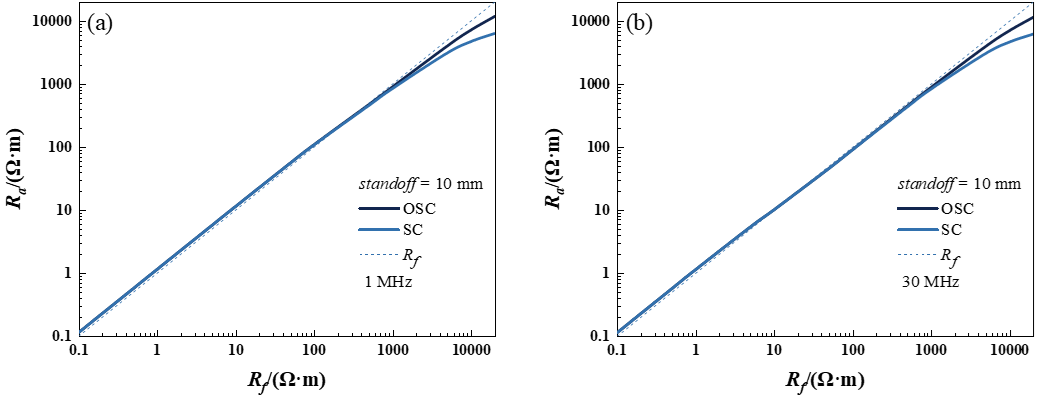}
    \caption{Comparison of open-short calibration and short calibration results ($R_m$ = 10000 $\Omega \cdot \mathrm{m}$, $\varepsilon_m$ = 8, $stdf$ = 10 mm). (a) Corrected Result at 1 MHz. (b) Corrected Result at 30 MHz.}
    \label{fig:9}
\end{figure}

To validate the correction effect of the OSC method on the standoff effect, the vertical coupling apparent resistivity, parallel coupling apparent resistivity \cite{7}, and the apparent resistivity calculated by the OSC method were compared and analyzed under the excitation frequencies of 1 MHz and 30 MHz (Figs.\ref{fig:6}-\ref{fig:9}).

Fig.\ref{fig:6} shows the apparent resistivity results calculated by the three methods under the conditions of mud resistivity 10000 $\Omega \cdot \mathrm{m}$ , mud relative permittivity 8, standoff 1 mm, and formation relative permittivity 20. It can be seen that the apparent resistivity calibrated by OSC has good continuity, which is basically consistent with the spliced vertical/parallel coupling apparent resistivity.

Fig.\ref{fig:7} and Fig.\ref{fig:8} compare the apparent resistivity calculation results under different formation permittivities and different standoffs, respectively. The analysis shows that when the formation relative permittivity is less than 100, the apparent resistivity calculated by the OSC method changes approximately linearly; when it is greater than 100, although the growth rate of apparent resistivity in high-resistivity formations slows down, the apparent resistivity can still be effectively calculated. In contrast, the calculable range of the vertical/parallel coupling method shrinks significantly with the increase of formation permittivity; the standoff has little effect on the OSC method, while the calculable range of the vertical/parallel coupling method decreases with the increase of standoff.

In general, the OSC method has a better correction effect on the impedance contribution of the mud layer than the vertical/parallel coupling method, and has a wider range of correctable formation resistivity. Moreover, a complete apparent resistivity curve can be obtained without splicing, which significantly reduces the uncertainty of the results.

Fig.\ref{fig:9} compares the apparent resistivity results of open-circuit-short-circuit combined calibration and short-circuit-only calibration. It can be seen that there is a certain difference between the two only in ultra-high resistivity formations with large standoff; in most cases, a high-precision apparent resistivity curve can be obtained by using only short-circuit calibration.

\subsection{Validation of the objective function}

\begin{figure}[!htbp]
    \centering
    \includegraphics[width=0.8\linewidth]{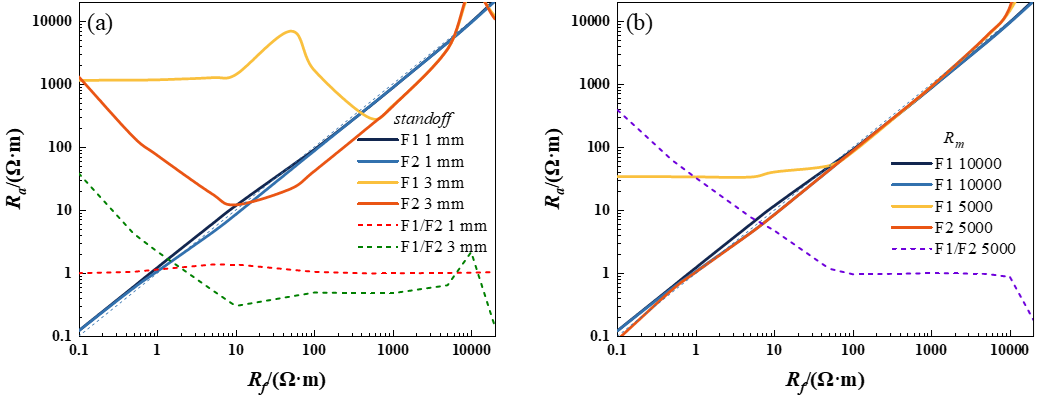}
    \caption{Validation of the resistivity consistency}
    \label{fig:10}
\end{figure}

The mud resistivity was set to 10000 $\Omega \cdot \mathrm{m}$ , the mud relative permittivity was set to 6, and the standoff was set to 1 mm. Fig.\ref{fig:10} presents the apparent resistivity calculation results under different deviated mud parameters, where Fig.\ref{fig:10}(a) shows the case of deviated standoff from the true value, and Fig.\ref{fig:10}(b) shows that of deviated mud resistivity. As can be seen from the figure: when the mud parameters are properly selected, the apparent resistivities at the two frequencies exhibit the best consistency, and their ratio is closest to 1; when the mud parameters are improperly selected, the apparent resistivities become discrete, with significant differences between the results at the two frequencies; the mud resistivity has a relatively small impact on the apparent resistivity calculation results. Even if its value is inaccurate, the calculated apparent resistivities are still very close to the true values within a certain range of formation resistivity.

\begin{figure}[!htbp]
    \centering
    \includegraphics[width=0.7\linewidth]{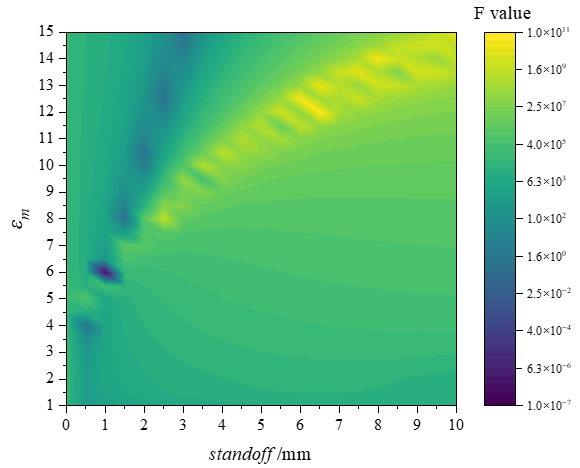}
    \caption{The distribution of objective function values}
    \label{fig:11}
\end{figure}

As shown in Fig.~\ref{fig:11}, the distribution of objective function values under different mud relative permittivities and standoffs is presented. The minimum objective function value occurs when the mud relative permittivity $\varepsilon_{m} = 6$ and the standoff is 1, which further validates the feasibility of the resistivity-consistency-based iterative inversion method.

\begin{figure}[!htbp]
    \centering
    \includegraphics[width=0.6\linewidth]{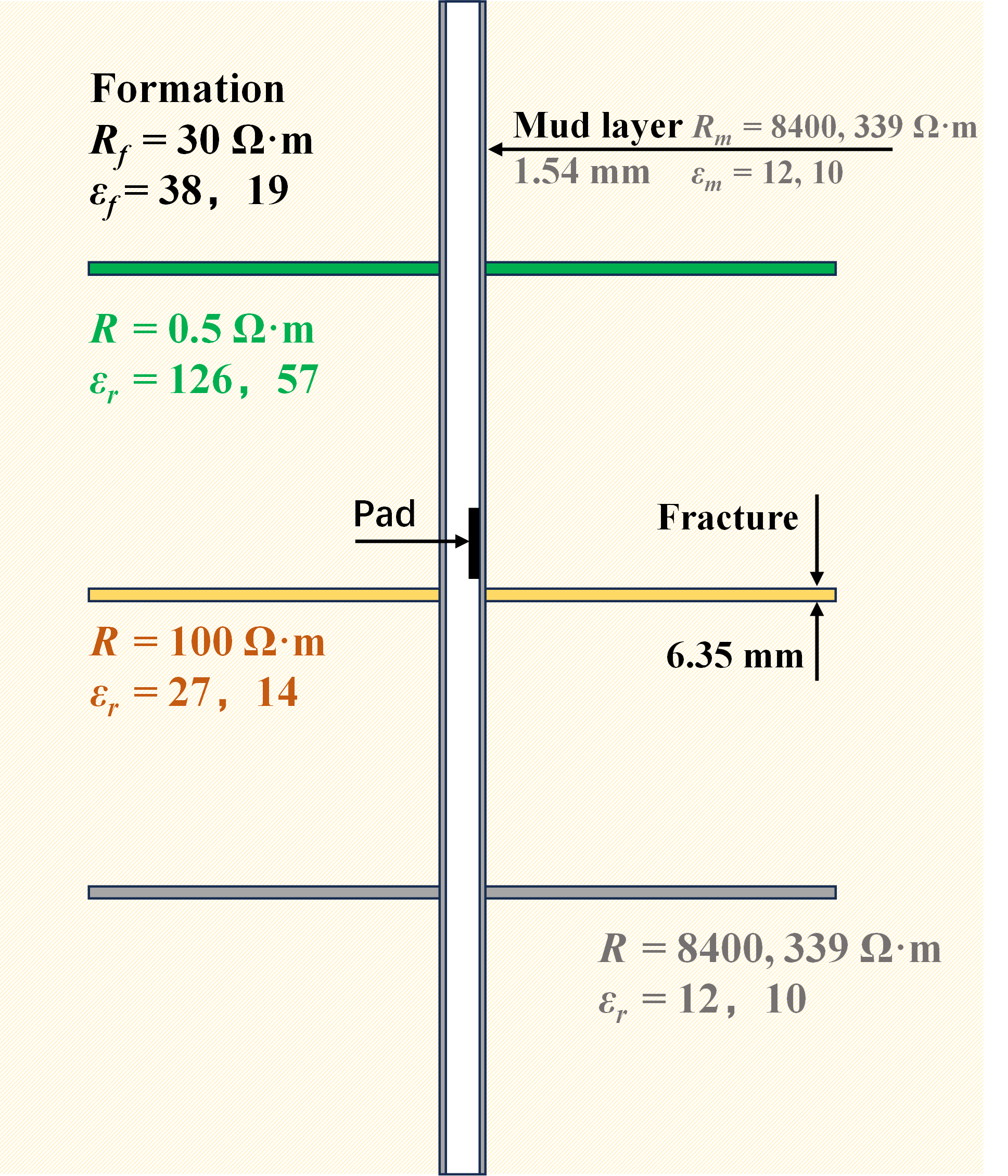}
    \caption{The fractured formation models}
    \label{fig:12}
\end{figure}

\begin{figure}[!htbp]
    \centering
    \includegraphics[width=0.8\linewidth]{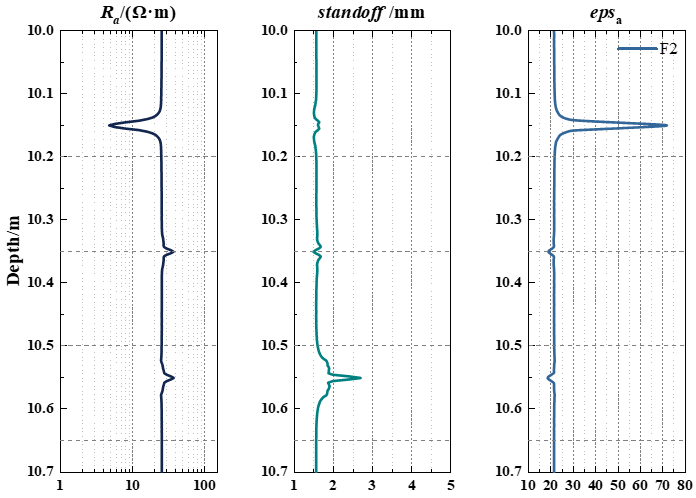}
    \caption{Resistivity consistency inversion results for fractured formations}
    \label{fig:13}
\end{figure}

The Gauss-Newton algorithm was used to invert the formation model containing three fractures filled with different materials. The formation model parameters were referenced from Ref.\cite{8} (Fig.\ref{fig:12}): standoff was 1.54 mm; at frequency 1, the mud resistivity was 8400 $\Omega \cdot \mathrm{m}$  and the relative permittivity was 12; at frequency 2, the mud resistivity was 339 $\Omega \cdot \mathrm{m}$  and the relative permittivity was 10; the formation resistivity was 30 $\Omega \cdot \mathrm{m}$ , and the relative permittivities at the two frequencies were 38 and 19, respectively; the fracture spacing was set to 20 mm, the fracture apertures were all 6.35 mm, and the fracture resistivities increased sequentially. The inversion results are shown in Fig.\ref{fig:13}, including the curves of apparent resistivity, mud thickness and relative permittivity at high frequency. From the inversion results, it can be concluded that: for the formation model with frequency dispersion characteristics, the consistency iterative inversion method still maintains good stability, and the inverted parameters can clearly reflect the changes in formation physical properties; the apparent resistivity curve is more sensitive to low-resistivity fractures; in high-resistivity fractures, the response of the standoff curve is more obvious, especially when the fractures are filled with mud; the variation trend of the relative permittivity curve at frequency 2 is opposite to that of the resistivity curve.

\FloatBarrier 
\section{Conclusions}

We introduce the open-short calibration method into the measurement signal processing of the oil-based mud resistivity imagers, and propose a new mud impedance calculation method. The measurement system is modeled as a three-layer model including the impedances of the electrode, mud layer, and external medium. The finite element method is used to simulate the impedance responses under open-circuit and short-circuit states, and the calibration formula is applied to eliminate the signals from the system and mud, thus effectively separating the formation signal. Validations on multiple models show that this method is superior to the vertical/parallel coupling method. It can realize full-range resistivity calculation from low-resistivity to high-resistivity formations and avoid the uncertainty caused by resistivity threshold splicing.

Taking the multi-frequency resistivity consistency residual as the objective function, we propose an approximate apparent resistivity consistency iterative inversion method. Only a small number of datasets need to be pre-constructed for this method, which can complete the optimal solution of 5 parameters (mud layer parameters) through 4 measured values, realizing the inversion of formation apparent resistivity, permittivity, interval and other parameters. The calculation efficiency is 2~3 times higher than that of quantitative inversion. The analysis and verification of the defined objective function show that when the mud layer parameters are reasonably selected, the objective function attains the minimum value, indicating that the resistivity consistency residual can be used to invert the formation resistivity. The theoretical verification on the fractured formation model and the application test on the case well show that the consistency inversion can effectively calculate the formation characteristics. Although it is an approximate processing method, its inversion results are highly consistent with the quantitative inversion results.

\bibliographystyle{ieeetr}
\bibliography{ref}

\end{document}